# Search for Magnetic Excitation Spectra of $Na_xCoO_2 \cdot yD_2O$
## -Neutron Scattering-

Taketo Moyoshi, Yukio Yasui, Minoru Soda, Yoshiaki Kobayashi, Masatoshi Sato
and Kazuhisa Kakurai[1]
*Department of Physics, Division of Material Science, Nagoya University, Furo-cho, Chikusa-ku, Nagoya 464-8602*
[1]*JAEA, Ibaraki 319-1195*

Neutron inelastic scattering studies on aligned crystals of $Na_xCoO_2 \cdot yD_2O$ have been carried out. Significant scattering has been observed for phonons originating from the vibrations of the short-range ordered $D_2O$ molecules. No firm evidence for the magnetic fluctuations has been found for the $Na_xCoO_2 \cdot yD_2O$, while the very weak intensity modulation observed in the scans along (1/2, 1/2, $l$) suggests the existence of the antiferromagnetic fluctuations in the non-deuterated parts of the sample, the volume fraction of which is estimated to be about 26 %. The present results indicate that magnetic fluctuations, especially ferromagnetic ones are rather weak, in the deuterated system, which is consistent with the results of the NMR Knight shift reported by the present authors' group. We discuss the short-range ordered state of $D_2O$, too by analyzing the observed diffuse scattering.

## 1. Introduction

$Na_xCoO_2 \cdot yH_2O$ ($x \sim 0.3$ and $y \sim 1.3$) was found by Takada *et al.*[1] to exhibit the superconducting transition at temperature $T = T_c \sim 4.5$ K. The system can be obtained from $Na_xCoO_2$ by de-intercalating $Na^+$ ions and then, by intercalating $H_2O$ molecules. The superconductivity has attracted much attention, because the system has the 2-dimensional triangular lattice formed by the $CoO_2$-layer stacking, in which electrons are expected to be strongly correlated and frustrated. Up to now, various symmetries of the order parameter have been theoretically proposed.[2-7]

Experimental studies have also been carried out intensively for both hydrated and non-hydrated systems. Among these studies, NMR measurements on polycrystalline and/or single-crystal samples have been applied to investigate the symmetry of the order parameter[8-13] and specific-heat ($C$) measurements have also been carried out.[14-16] Although the possibility of the triplet pairing has been proposed by many authors, the suppression of the spin component of the NMR Knight shift observed below $T_c$ by the present authors' group for both field directions within and perpendicular to the $CoO_2$ planes, indicates that the Cooper pairs are in the singlet state. Rather small rate of $T_c$-suppression by non-magnetic impurities indicates that the symmetry seems to be *s*-like.[17]

On the non-hydrated system $Na_xCoO_2$, measurements of the magnetic susceptibilities and specific heats have revealed characteristic $x$ dependence of the electronic nature of $Na_xCoO_2$.[18,19] Neutron structural studies on powder samples[20,21] and the angle-resolved photoemission spectroscopy (ARPES)[22,23] have also been carried out to clarify the structural and electronic characteristics. On the magnetic structures in the ordered phase found in the region of $x \geq 0.7$, Boothroyd *et al.*,[24] Bayrakci *et al.*[25] and Helme *et al.*[26,27] have pointed out by neutron scattering measurements on single-crystal samples of $Na_xCoO_2$ that the spins align ferromagnetically within a $CoO_2$ plane with their directions perpendicular to the plane. They also discussed the magnetic exchange coupling constants. For $x = 0.5$, the present authors' group[19] proposed possible magnetic structures based on their data of neutron diffraction and NMR studies: Two crystallographically distinct Co sites carry different magnetic moments. The larger moments lie within the plane and the in-plane arrangement is antiferromagnetic, while the directions of the smaller moments are perpendicular to the planes. Gašparović et al.[28] have also studied the magnetic structure by using polarized neutrons. These results indicate that the in-plane magnetic interaction is ferromagnetic in the Na rich region and antiferromagnetic in the Na poor region.

---

*corresponding author: (e-mail: e43247a@nucc.cc.nagoya-u.ac.jp)

In order to see whether magnetic fluctuations play an important role for the realization of the superconductivity, we have searched for magnetic excitations by using aligned crystals of deuterated samples with the total volume of ~2 cm$^3$. In the experiment, we have observed rather strong diffuse scattering, which can be understood as the contribution from the intercalated $D_2O$ molecules.

## 2. Experiments

Single crystals of $Na_xCoO_2$ with $x \sim 0.72$ were grown by the floating zone (FZ) method, as reported in ref. 29. These crystals were immersed into the $Br_2/CH_3CN$ solution for several days and washed with $CH_3CN$. Then, the $D_2O$ intercalation was carried out. After the intercalation, the cross section parallel to the ab plane of the crystal changed to an elliptical one with the longer and shorter axes being ~6 mm and ~3 mm, respectively. Three crystals were aligned and used in the neutron measurements.

Figures 1(a) and 1(b) show the magnetic susceptibilities χ measured by SQUID magnetometer for parts of the crystals used in the present study. The superconducting transition can be seen just above 4 K. Neutron measurements were carried out by using the triple axis spectrometers TAS-1 installed at the JRR-3M of JAEA in Tokai. The horizontal collimations were 40'-40'-80'-80'. The sample was set in an Al can filled with He exchange gas. The crystals were with the [001] and [110] axes in the scattering plane. The temperature was controlled by using a Displex-type refrigerator. The 002 reflections of Pyrolytic graphites (PG) were used for both the monochromator and the analyzer. Scattered neutron energy ($E_f$) was fixed at 14.7 meV. A PG filter was placed after the sample to eliminate the higher order contamination. To avoid effects of the neutron absorption, the aligned-crystals were set with their rod-axes nearly vertical. (We could choose such the setting, because the [001] axis and one of the [110] axes are about perpendicular to the rod axis.)

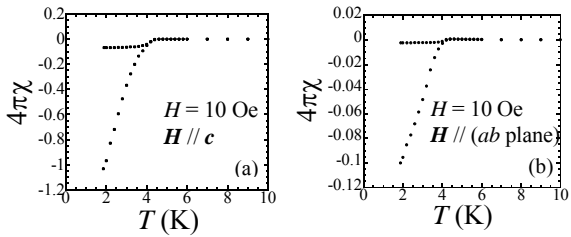

Fig. 1 Magnetic susceptibilities of single-crystal samples of $Na_xCoO_2 \cdot yD_2O$, measured under the magnetic fields **H** // **c** (a) and **H** // (**ab**-plane) (b).

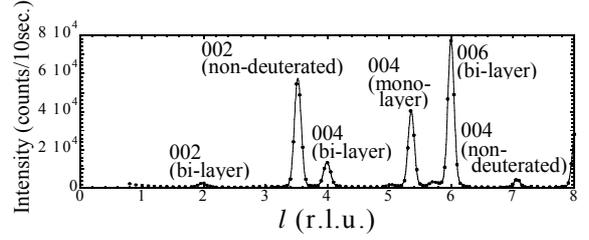

Fig. 2 Neutron elastic scattering intensity measured along (0, 0, l) at 5.5 K for the aligned crystals of $Na_xCoO_2 \cdot yD_2O$. Peaks can be indexed by the reflections of $Na_xCoO_2$ or $Na_xCoO_2 \cdot yD_2O$ with y ~ 2/3(mono-layer) or 4/3(bi-layer).

Figure 2 shows the result of the elastic scan along (0, 0, l) taken at 5.5 K. Hereafter, indices for $Na_xCoO_2 \cdot yD_2O$ (x ~ 0.3 and y ~ 1.3; bi-layer system) are used, if no indication is given. In the figure, we can recognize Bragg reflections from parts of the superconducting $Na_xCoO_2 \cdot yD_2O$ (y ~ 1.3; bi-layer system), non-superconducting $Na_xCoO_2 \cdot yD_2O$ (y ~ 0.66; mono-layer one) and non-deuterated $Na_xCoO_2$. This indicates that the $D_2O$ intercalation was not perfect. By analyzing the intensities of these reflections, the volume ratios of bi-layer, mono-layer and non-deuterated systems were estimated to be 64 %, 10 % and 26 %, respectively. (In the analysis, we used the structures reported by Argyriou et al.[30], Sakurai et al.[31] and Huang et al.[20] for these systems, respectively.) The Na concentration $x$ of non-deuterated parts is estimated to be ~0.44 from the $c$ value.[18]

## 3. Experimental Results and Discussion

Figure 3 shows the result of constant-$\Delta E$ scan taken at 5.5 K along (1/2, 1/2, l) with the transfer energy $\Delta E$ = 2.5 meV. The dashed line is a sinusoidal one drawn by fitting to the data. We can recognize very weak modulation, which has a periodicity of the reciprocal lattice vector **c\*** of non-deuterated parts. It becomes maximum at points with odd $l$'s and minimum at even $l$'s in the reciprocal space of non-deuterated parts, suggesting that the magnetic correlation in the non-deuterated parts is antiferromagnetic both in-plane and inter-plane directions. It is consistent with the arguments by the present authors' group that there is the antiferromagnetic correlation in the region $x \leq 0.6$ in $Na_xCoO_2$. We have also tried the similar scan with the smaller transfer energy $\Delta E$ = 2.0 meV, expecting the stronger intensity modulation. However, as shown in Fig. 4, there exists a significant elastic diffuse component along (1/2, 1/2, $l$) and it contaminates the modulation as a result of finite resolution in the region of small $l$.

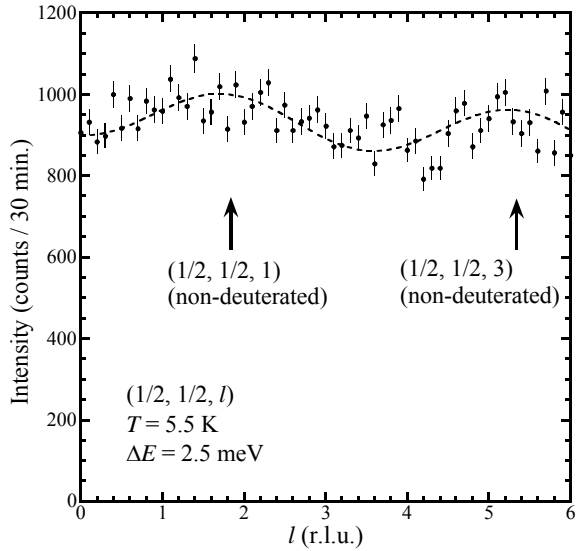

Fig. 3  Neutron scattering intensity measured along (1/2, 1/2, $l$) for $\Delta E$ = 2.5 meV at 5.5 K. The dashed line is the fitted sinusoidal curve with a linear background counts. Indices by using the lattice parameters of the non-deuterated system are indicated at the corresponding positions.

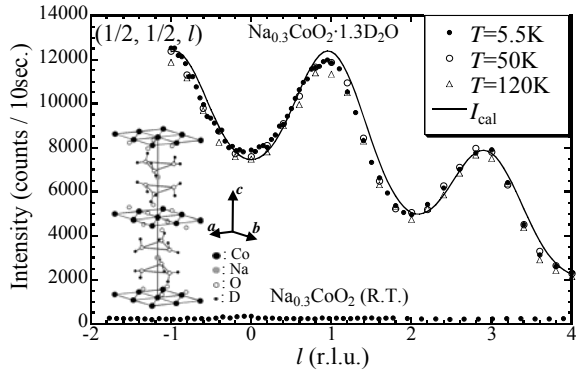

Fig. 4  Neutron elastic scattering intensity measured along (1/2, 1/2, $l$) at 5.5 K, 50 K and 120 K on the aligned crystal samples of $Na_xCoO_2 \cdot yD_2O$. The data taken at room temperature on aligned crystals of $Na_{0.3}CoO_2$ are also shown, where the abscissa shows the scattering vector $Q$ in unit of $c^*$ of bi-layer system. The solid line is a result of a model calculation (see text for details). Inset shows the unit cell of the crystal structure proposed by Argyriou et al. The solid line along $c$ connects the center of gravities of the triangles formed of three oxygens of $D_2O$ molecules.

The diffuse component has an intensity modulation with a periodicity of $c^*$ of the bi-layer system. Because the intensities do not change with increasing temperature, the scattering is considered not to be due to magnetic origins but due to the atomic or molecular ordering, which may correspond to the $D_2O$ ordering observed by Argyriou et al.[30] in their electron and neutron diffraction studies. Discussion on this diffuse component is presented later and here, we just note that the ordering is short-ranged along the $c$ direction, while the in-plane correlation of $D_2O$ or the correlation length within a layer is almost infinite (at least more than 300 Å), judging from the widths of the 1/21/20 reflection shown in the inset of Fig. 5.

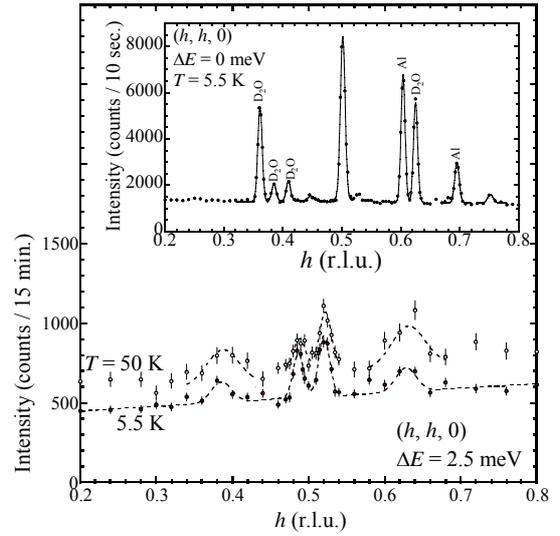

Fig. 5  Neutron scattering intensity measured along ($h$, $h$, 0) for $\Delta E$ = 2.5 meV at 5.5 K and 50 K. The inset shows the result of elastic scan along ($h$, $h$, 0) at 5.5 K, where the Bragg reflections of $D_2O$ powder and Al are observed together with the diffuse scattering at $h$ = 1/2, which spreads widely along the $c$ direction.

Results of $h$-scans along ($h$, $h$, 0) are shown in Fig. 5 for $\Delta E$ = 2.5 meV at 5.5 K and 50 K. We find two peaks around $h$ = 1/2 and these peaks spread widely along $l$, as the elastic component shown in Fig. 4. We think that they are from phonons, because the single elastic peak observed at $h$ = 1/2 splits into the two, when $\Delta E$ becomes nonzero, suggesting that the excitation has the dispersion. These phonons probably related to the intercalated $D_2O$ molecules, which order almost 2 dimensionally as stated above. Peaks are also found at around $h$ = 0.385 and 0.63. We consider them to be inelastic and from the powder of $D_2O$ ice. (The observation of two Bragg peaks shown in the inset of Fig. 5 is one of the evidence for the existence of the Debye ring of $D_2O$.)

The constant-$\Delta E$ scans along (0, 0, $l$) at 5.5 K

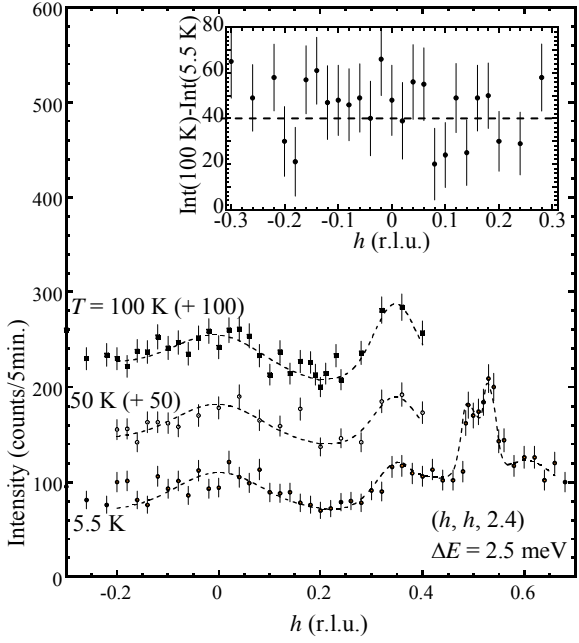

Fig. 6  Neutron scattering intensities measured along $(h, h, 2.4)$ for $\Delta E = 2.5$ meV at 5.5 K, 50 K and 100 K. The dashed lines are the guides for the eye. The inset shows the difference between the data of 5.5 K and 100 K, which shows that only the background counts changes with temperature.

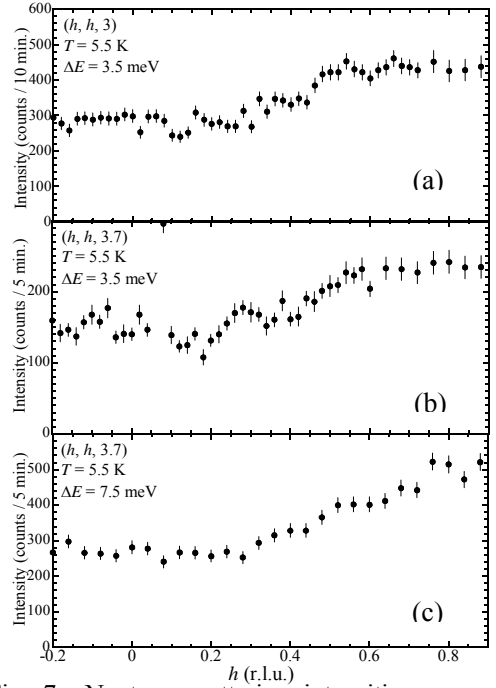

Fig. 7  Neutron scattering intensities measured along $(h, h, 3)$ at $\Delta E = 3.5$ meV (a) and along $(h, h, 3.7)$ at $\Delta E = 3.5$ meV (b) and 7.5 meV (c). $T = 5.5$ K.

with $\Delta E = 2.5$ meV have also been carried out to study if there exist in-plane ferromagnetic fluctuations. However, we have not obtained meaningful results, or the signal is less than the experimental error bars determined by the small but non-negligible contaminations caused by the effect of the direct beam in the region of small $l$ and due to the $00l$ Bragg reflections in the region of the relatively large $l$.

Then, we have carried out $h$-scans along $(h, h, l)$ for several fixed $l$'s at three temperatures. (We assume that the magnetic correlation in the bi-layer system is, if it exists, 2-dimensional, that is, its contribution to the scattering is extended over the wide $l$ region.) Figure 6 shows the result along $(h, h, l = 2.4)$, where the $l$ value corresponds to the point where the background count becomes minimum in the scan along $(0, 0, l)$. In the figure, four peaks are observed at $h = 0$, 0.35, 0.48 and 0.52. One might think that the peak at $h = 0$ corresponds to the in-plane ferromagnetic fluctuations. However, the intensity of this peak remains constant with temperature $T$ at least up to 100 K (except the increase of the background counts, as shown in the inset by plotting the difference between the count numbers at 5.5 K and 100 K against $h$). It indicates that the peak is not from the magnetic origin. (Actually, the peak originates from the direct beam, because the scattering angle is rather small. We can confirm it by the fact that the peak disappears in the scans with the larger $l$, as can be seen in Fig. 7.) Other peaks at $h = 0.35$, 0.48 and 0.52 correspond to those observed for $l = 0$ at nearly the equal values of $h$, respectively. The absolute value of the scattering vector (0.35, 0.35, 2.4) is found to be equal to that of (0.385, 0.385, 0), the peak position shown in Fig. 5, supporting the idea that they are related to the Debye ring of $D_2O$ ice.

We have carried out $h$-scans along $(h, h, l)$ with different values of $l$ and $\Delta E$. Figures 7(a)–7(c) show results of $h$-scans along $(h, h, 3)$ and $(h, h, 3.7)$ at 5.5 K. $\Delta E = 3.5$ meV for $l = 3$ and they are 3.5 and 7.5 meV for $l = 3.7$. Around $h = 0$, the intensities are approximately flat in the three scans. Then, even if there exist ferromagnetic fluctuations, it is not in the $\Delta E$ region up to 7.5 meV. In Figs. 7(a)–7(c), the observed intensity increases gradually with increasing $h$ in the region $h > 0.3$ and seems to have a broad peak at around $h = 1/2$, which may support the existence of the antiferromagnetic fluctuations in the present aligned crystals.

Now, we have presented the data taken in the search for the magnetic excitations, which many researchers expect, at first sight, to be important for the occurrence of the superconductivity. However, we have obtained no firm evidence for the existence of the magnetic excitations. Instead, the excitations related to the vibration of $D_2O$ molecules have been observed. The intensity-modulation shown in Fig. 3 suggests, though it is very weak, the existence of the 2-dimensional antiferromagnetic fluctuations in non-deuterated system. Because the deuterated superconductor has the 1.4 times larger molar fraction than the non-deuterated system in the present specimen, the stronger intensity have to be observed in the former system, if the similar amounts of fluctuations exist in these systems. However, at least, the fluctuations, which are ferromagnetic within a plane have not been observed.

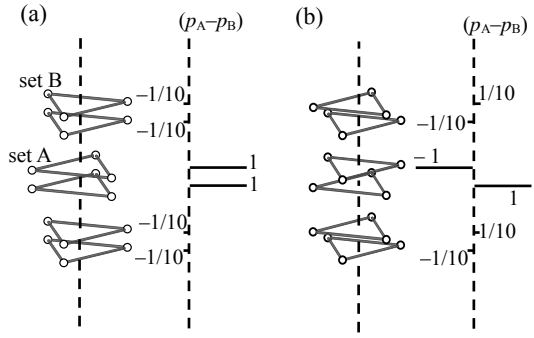

Fig. 8 Model of the short range correlation of $D_2O$ along $c$, (a) (AA)(BB)(AA) and (b) (AB)(BA)(AB), is shown by using the order parameter $\phi$ ($\equiv p_A - p_B$). The domains of the patterns (a) and (b) have equal weights. The in-plane correlation is rather long ranged (see text for details).

Finally, we present the result of the analysis of diffuse intensity. To clarify the origin of the observed diffuse scattering, we consider the structure proposed by Argyriou et al..[30] According to them, three $D_2O$ molecules form a regular triangle isolated from each other within a $D_2O$ layer, as shown in the inset of Fig. 4. The centers of gravity of these triangles stack straight along the $c$ axis, forming $D_2O$ bi-layer structure. The distance between the bi-layers is $c/2$. The triangles can be divided into two sets A and B: The arrangement of a triangle of one set (say set A) with respect to the crystal axes $a$ and $b$, can be obtained by rotating a triangle of another set (set B) by 60 degrees about the $c$ axis. The Bragg reflections appear at odd $l$ values, if two kinds of triangle-stacking along the $c$ direction ···(AA)(BB)(AA)··· and/or ···(AB)(BA)(AB)··· exist, where the parentheses indicate the bi-layers. Considering that the data shown in Fig. 4 definitely indicate the diffuse nature of the reflection, we employ, for an explanation of the observed data, a simple model of short-range correlation of six $D_2O$ layers (three bi-layers) along the $c$ axis: Describing the probability of the triangular arrangements, set A and set B as $p_A$ and $p_B$ ($p_A + p_B = 1$), the spatial correlation of the order parameter $\phi \equiv p_A - p_B$ is assumed to have a form shown in Fig. 8, where the domains with two different triangle-stackings exist. Using the model and with the consideration of the resolution volume, we can reproduce the observed data as shown in Fig. 4 by the solid line. We note here that not only the $Q$ dependence of the intensity of the scattering but also the intensity itself relative to the Bragg intensity can roughly be reproduced. The present result clearly indicates that the observed diffuse scattering originates from the short-range ordered $D_2O$ molecules. However, we do not know why the two different stacking sequences ···(AA)(BB)(AA)··· and ···(AB)(BA)(AB)··· appear with the equal weight.

## 4. Summary and Conclusion

We have performed neutron inelastic scattering studies on aligned crystals of $Na_xCoO_2 \cdot yD_2O$ with the total weight of ~5 g. Although the significant intensity has been observed for the phonons, originating from the vibrations of the short-range ordered $D_2O$ molecules, we have not found any firm evidence for the magnetic fluctuations in the deuterated superconductor $Na_xCoO_2 \cdot yD_2O$, while the very weak intensity modulation observed in the scans along (1/2, 1/2, $l$) (Fig. 3) suggests the existence of the 2-dimensional antiferromagnetic fluctuations in the non-deuterated system.

The present results indicate that magnetic fluctuations are rather weak, if they exist. It can be said especially for ferromagnetic fluctuations and is consistent with the results of the NMR Knight shift reported by the present authors' group.[11] In this sense, it is interesting to note that there has been pointed out that the superconductivity can be naturally explained by considering the electron-phonon coupling.[7] For a much more quantitative estimation of the strength of the magnetic fluctuations in $Na_xCoO_2 \cdot yD_2O$, further studies have to be carried out. Finally, we have discussed the short-range ordered state of $D_2O$ by observing the significant diffuse scattering.

**Acknowledgements**

The authors thank Mr. K. Yada and Prof. H. Kontani for fruitful discussion. Work at JRR-3M was performed within the frame of JAEA Collaborative Research Program on Neutron Scattering. The work is supported by Grants-in-Aid for Scientific Research from the Japan Society for the Promotion of Science (JSPS) and by Grants-in-Aid on priority area from the Ministry of Education, Culture, Sports, Science and Technology.